\documentclass[preprint,12pt,authoryear,psfig,linenumbers,showpacs,superscriptaddress,reprint,floatfix]{elsarticle}
\usepackage{amsmath}
\usepackage{mathtools}
\usepackage{amssymb}
\usepackage{color}
\usepackage{graphics}
\usepackage{epsfig}
\usepackage[normalem]{ulem} 
\usepackage{cancel}
\usepackage[mathlines]{lineno}
\graphicspath{{Figures/}}
\usepackage{xtab,afterpage,longtable}
\usepackage{booktabs}   
\usepackage{ltablex}
\usepackage{float}
\begin{document}
\begin{frontmatter}


\title{Hydrodynamic energy flux in a many-particle system} 
\author[Jammu]{Rauoof Wani\corref{cor1}}
\ead{abrauoofwani@gmail.com}
\cortext[cor1]{ Corresponding author: Rauoof Wani}
\author[Kanpur]{Mahendra Verma}
\author[Kanpur]{Shashwat Nirgudkar}
\author[Jammu]{Sanat Tiwari}
\affiliation[Jammu]{organization={Indian Institute of Technology Jammu, Department of Physics},
            city={Jammu},
            postcode={181221}, 
            country={India}}

\affiliation[Kanpur]{organization={Indian Institute of Technology Kanpur, Department of Physics},
            city={ Kanpur},
            postcode={208016},
            country={India}}

\begin{abstract}
In this letter, we demonstrate energy transfers and thermalization in an isolated ensemble of realistic gas particles. We perform a grid-free classical molecular dynamics simulation of two-dimensional Lenard-Jones gas. We start our simulation with a large-scale vortex akin to a hydrodynamic flow and study its non-equilibrium behavior till it attains thermal equilibrium. In the intermediate phases, small wavenumbers ($k$) exhibit $E(k) \propto k^{-3}$ kinetic energy spectrum, whereas large wavenumbers exhibit $E(k) \propto k$ spectrum. Asymptotically, $E(k) \propto k$ for the whole range of $k$, thus indicating thermalization. These results are akin to those of  Euler turbulence despite complex collisions and interactions among the particles.
\end{abstract}
\end{frontmatter}

\section{introduction}
Turbulent flows exhibit multi-scale energy cascade, which has been derived using Navier-Stokes and Euler equations~\cite{Kolmogorov:DANS1941Structure,Kolmogorov:DANS1941Dissipation,Frisch:book}. However, a hydrodynamic fluid is a collection of random molecules. Hence, it is imperative to derive energy transfers in a multi-particle system. In this letter, we capture such energy transfer in a multi-particle system using high-resolution molecular dynamics (MD) in two dimensions (2D). We simulate a large hydrodynamic vortex superposed on a thermal background in an isolated system with total energy conserved. The vortex in the system generates an energy cascade from large to small scales, akin to that in hydrodynamic turbulence~\cite{Kolmogorov:DANS1941Dissipation}.

According to kinetic theory ~\citep{bellomo1997boltzmann} and statistical mechanics~\citep{huang2008statistical,Reif:book:StatMech,goldstein2001boltzmann}, multiparticle systems thermalize or reach equilibrium after a sufficiently long time,  whose behavior is described by thermodynamic laws~\citep{Grad1958,loeb2004kinetic}. A popular multiparticle system is Lenard-Jones (LJ) gas, where particles interact through the short-range LJ potential. In this letter, we demonstrate that an LJ gas with vortex exhibits energy flux and thermalization after the hydrodynamic energy is fully converted to thermal energy. This scenario resembles thermalization in three-dimensional (3D) Euler turbulence~\cite{Cichowlas_PRL_2005}. This is the first ab initio or first-principle MD simulations exhibiting turbulent energy flux.

Euler turbulence is studied using an inviscid, incompressible fluid equation given as:
\begin{equation}
    {\partial_t} {\bf u} +  {\bf u \cdot \nabla u} = -\nabla p;~~~{\bf \nabla \cdot  u} = 0, 
\label{eq:Euler_eqn}    
\end{equation}
where ${\bf u}, p$ are the velocity and pressure fields respectively.  An Euler flow conserves total kinetic energy ($\int d{\bf x} u^2/2$). In addition, 2D and 3D Euler equations conserve, respectively, the total enstrophy ($\int d{\bf x} \omega^2/2$)  and total kinetic helicity ($\int d{\bf x} ({\bf u} \cdot \boldsymbol{\omega}) $), where 
$\boldsymbol{\omega} = \nabla \times {\bf u}$. Lee \citep{Lee:QAM1952} and Kraichnan \citep{Kraichnan:JFM1973} showed that Euler turbulence admits equilibrium solutions that exhibit approximately $k^2$ energy spectrum for 3D and $k$ energy spectrum for 2D (also see \cite{Onsagar:Nouvo1949_SH}). For the 3D Euler turbulence, Cichowlas $\textit{et al.}$
\citep{Cichowlas_PRL_2005} showed the Taylor-Green vortex evolves from $k^{-5/3}$ spectrum with positive energy flux at the small and intermediate wavenumbers and  $k^2$ spectrum with zero energy flux at large wavenumbers. Eventually, the system thermalizes and exhibits $k^2$ spectrum for all wavenumbers. Note that $k^{-5/3}$ spectrum and constant energy flux are the predictions by Kolmogorov~\cite{Kolmogorov:DANS1941Structure,Kolmogorov:DANS1941Dissipation,Frisch:book} for fully-developed hydrodynamic turbulence.
\\
For two-dimensional Euler turbulence, some researchers advocate it reaching equilibrium asymptotically  ~\citep{robert_sommeria_1991,kraichnan_1973_JOFM,Lee:QAM1952,Onsagar:Nouvo1949_SH}, whereas others argue it to be out of equilibrium in the final state~\citep{Pakter:PRL2018,Bouchet:PRL2009,Seyler:PF1975,Verma:PRF2022}. Further,  the initial condition of 2D Euler turbulence affects the asymptotic states considerably; in some cases, the asymptotic states are in equilibrium, but in many cases, they are out of equilibrium. \citep{Bouchet:PRL2009,Bouchet:PR2012,Verma:arxiv2020_equilibrium,Verma:PRF2022}. 
\\
Now, turning to particle simulations, many papers and books describe various properties of fluids in equilibrium or near-equilibrium using the Lattice Boltzmann Method (LBM),  Direct Simulation Monte Carlo (DSMC), and Molecular Dynamics (MD). But, in this letter, we focus on the connections between turbulence and thermal particles. In past LBM and DSMC simulations,
 Gallis $\textit{et al.}$, McMullen $\textit{et al.}$, Martinez $\textit{et al.}$, Chen and Doolen, Bird
\citep{Martinez:PF1994,chen1998lattice,bird1976molecular,Gallis:PRL2017, McMullen_2022_PRL} capture the physics at a granular resolution, in particular thermal fluctuations and hydrodynamic features, e.g., instabilities~\citep{Kadau_PNAS_2004, Ashwin_PRL_2015, Wani_2022}, waves \citep{POP_2022_SKT}. In particular, Gallis $\textit{et al.}$ \citep{Gallis:PRL2017} and McMullen $\textit{et al.}$ \citep{McMullen_2022_PRL} simulated 3D decaying hydrodynamic turbulence using DSMC  and showed that the resulting spectrum is a combination of $k^{-5/3}$ and exponential. Here, the energy of the structures cascades to small scales and finally to the thermal part,  akin to those for 3D Euler turbulence. 
\\
Several important works attempt to understand fluid's macroscopic and microscopic features simultaneously by coupling the Navier-Stokes equation to fluctuating hydrodynamics. In this framework,  Bell $\textit{et al.}$ \citep{Bell:JFM2022} and Bandak $\textit{et al.}$\citep{Bandak:PRE2022} showed that such systems exhibit Kolmogorov's $k^{-5/3}$ and  $\exp(-k \eta)$ ($\eta$ is the Kolmogorov's length scale) energy spectra in the inertial-dissipation range,  but it transitions to $k^2$ energy spectrum after the exponential part. The $k^2$ spectrum is attributed to the thermalization processes at the small scales. Betchov \citep{Betchov:JFM1975} reported a similar crossover to the $k^2$ spectrum in his experiment on the turbulent jet. We remark that the connections between hydrodynamics and kinetic theory shed light on thermalization, the arrow of time,  measure of entropy for hydrodynamics, etc.~\cite{Frisch:book, Bandak:PRE2022, Gallis:PRL2017, Verma:EPJB2019, Verma:PTRSA2020, Verma:PRF2022}.
\\
In this letter, we perform molecular dynamics (MD) simulations of a large number of weakly interacting particles in a 2D box. For the initial condition, we chose a large hydrodynamic vortex embedded in a background of randomly moving thermal particles. We isolate this system from the heat bath to conserve the total kinetic energy. Hence, our system is similar to a noisy Euler flow whose total energy is conserved. We observe that our system starts with a nonequilibrium energy transfer from large scales to small scales, and it thermalizes after tens of eddy turnover times. Our MD simulation captures the dynamics more accurately than the DSMC and represents real gas density and the intrinsic transport properties of the medium. Our work demonstrates  energy flux and thermalization  in an LJ gas  that includes realistic interactions, 
gas density, and intrinsic transport properties of the medium. 
\section{Methodology}
In this letter, we report the results of an MD simulation with $2.5\times 10^5$ particles using the open-source code  LAMMPS (Large-scale Atomic/Molecular Massively Parallel Simulator)~\citep{LAMMPS_plimpton}. We employ MPI-based parallel LAMMPS with a maximum of 400 CPU cores. The particles, confined in a periodic 2D (x-y geometry) square box,  interact with each other via pairwise LJ potential of the form $\phi_{ij}(r_{ij})=4\epsilon \left[\left(\sigma/r_{ij} \right)^{12} - \left(\sigma/r_{ij}\right)^6\right]$, 
where $r_{ij}$ is the separation between the particles, $\sigma$ is the distance at which $\phi_{ij}$ is zero, and $\epsilon$ is the depth of the potential. In this letter, we choose a small $\sigma$ and isolate the system from the heat bath. Also, the kinetic energy dominates the potential energy for our chosen temperature. 
\begin{figure*}[htb!]
    \centering
    \includegraphics[width=\columnwidth]{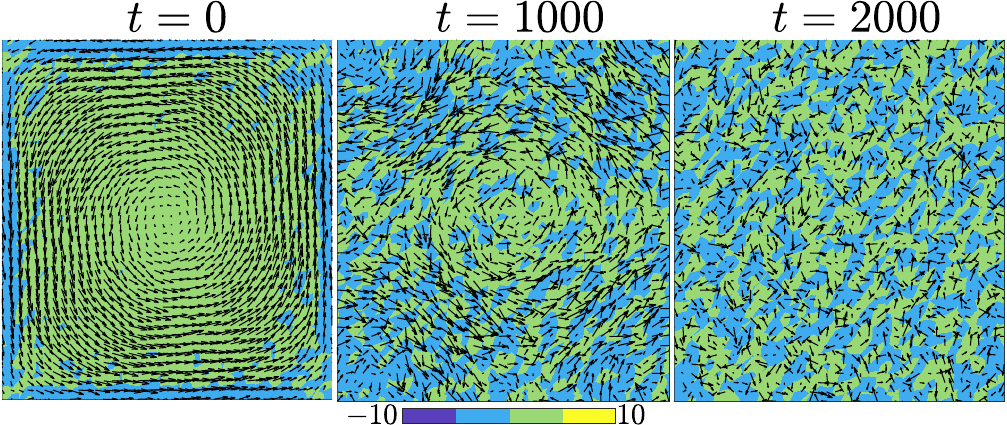}
    \caption{
    Snapshots of the velocity (represented by arrows) and vorticity (represented by colors) fields at $t=0, 1000$, and $2000$ LJ units in our MD simulation with a vortex structure embedded in a noisy environment as an initial condition. The system evolves from an ordered structure ($t=0$) to a thermalized state ($t=2000$).
    }
    \label{fig:TGV_evolution}
\end{figure*}

\begin{figure}[!ht]
    \centering
    \includegraphics[width=0.8\columnwidth]{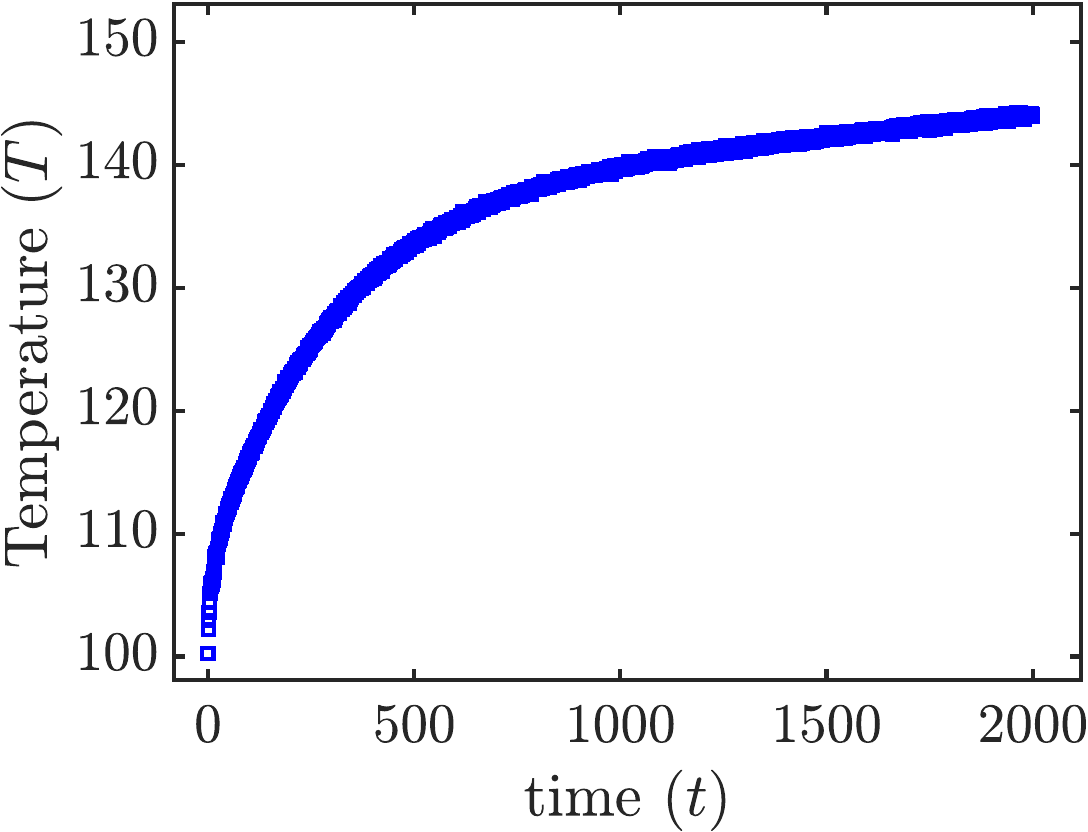}
    \caption{Temperature profile of LJ gas fluid up to $2000$ LJ-time units. The gradual increase in temperature with time is observed}
    \label{fig:fluid_temperature}
\end{figure}
In the present simulation, we employ the standard LJ units, in which $\sigma$ is the unit of length, and $\epsilon$ is the unit of energy. For our simulation, the box size $L$ is 500 LJ units, and the mean free path is $0.9$ LJ units, significantly shorter than the box size. Hence, our simulation satisfies the continuum approximation and effectively includes the impact of thermal fluctuations at hydrodynamic scales.
\\
We start with randomly moving particles  (white noise) whose rms value is $u_\mathrm{th} = \sqrt{2 T}$, where  $T$ is the temperature in LJ units (here, the particle mass $m = 1$, and the Boltzmann constant $k_B=1$). In addition, we provide each particle with a coherent velocity given by 
${\bf u(x},t=0) = A \left[\sin(k_0 x) \cos(k_0 y) \hat x - \cos(k_0 x) \sin(k_0 y) \hat y \right]$
 \label{eq:V_init_cond} 
 where $k_0= \pi/L$, and $ A=u_\mathrm{th}$ is a constant amplitude chosen so that the initial coherent and incoherent kinetic energies have similar magnitudes. Figure~\ref{fig:TGV_evolution} illustrates the initial condition and its evolution. The above initial condition mimics realistic terrestrial flows that have large-scale macroscopic velocity and small-scale thermal velocity. For present simulations, we take $T = 100 $ LJ units. Hence, rms velocity $U = \sqrt{200} \approx 14.1$ LJ units, and the eddy turnover time at $t=0$ is $L/U \approx 35.5$ LJ units. Our simulation can be related to inert Ar gas in a box of size 0.17 $\mu m$ whose average inter-particle separation is 0.19 nm and the associated mean free path is 0.304 nm. We also remark that our numerical results are independent of particle numbers if it is more than tens of thousands. 

After applying the vortex velocity profile, we let the LJ gas evolve from $t=0$ to $2000$ LJ units, which is approximately 56 eddy turnover time. In Figure \ref{fig:TGV_evolution}, we illustrate three snapshots of the flow using the velocity and vorticity fields at  $t = 0$, 1000, and 2000 LJ time units. These plots illustrate the evolution from an ordered state to a thermalized state, consistent with the numerical results of 3D Euler turbulence~\cite{,Cichowlas_PRL_2005,Verma:PRF2022}. We quantify the thermalization process using the kinetic energy spectrum and flux related to the hydrodynamic velocity computed on a 40x40 grid. For a given snapshot, we compute the coarse-grained flow velocity ${\bf u(x},t)$  by averaging over several hundred particles in each grid. We Fourier transform ${\bf u(x},t)$ and compute ${\bf u(k},t)$, and subsequently, the modal energy $E({\bf k},t) = |{\bf u(k},t)|^2/2$. After this, we compute the 1D shell spectrum $E(k,t) = \sum_{k-1 \le k' \le k} E({\bf k}',t)$ \citep{Frisch:book}. 
Our spectral plots in Fig.~\ref{fig:fig_power_spectrum}(a,b) exclude $E(k,t)$ at $k = {k_0, 2k_0}$. However, all the available wavevectors are considered for the computation of nonlinear transfer, energy fluxes, and enstrophy. 
\\
We also note that the temperature of the fluid gradually rises as time progresses, as shown in fig.~\ref{fig:fluid_temperature}. This is because the prominent flow patterns gradually decline over time, with the flow reaching thermal equilibrium approximately within 50 time units or one eddy turnover time. This behavior arises due to the low viscosity of the flow and a high Reynolds number, approximately 812. Consequently, the temperature of the flow undergoes a gradual increase.
\\
\subsection{Energy transfer analysis}
\begin{figure}[!ht]
    \centering
    \includegraphics[width=\columnwidth]{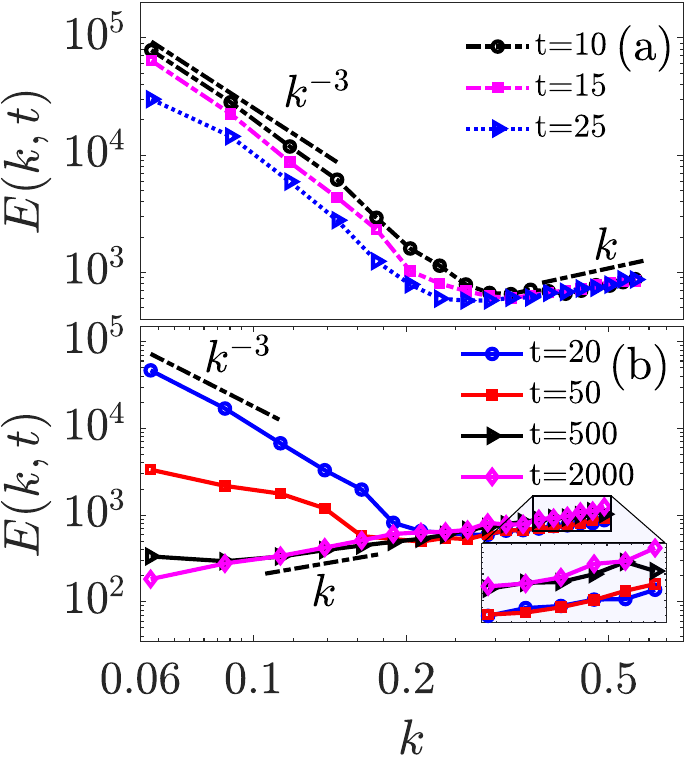}
   \caption{For the MD simulation, (a): Short-term evolution of the shell spectra $E(k,t)$ at $t=10, 15$, and 25 LJ time units. Here, $E(k,t) \propto k^{-3}$ at small wavenumbers, and $E(k,t) \propto k$ at large wavenumbers. (b): Long-term evolution of $E(k,t)$ at $t = 20, 50, 500$, and 2000 LJ time units. $E(k,t) \propto k$ for all $k$'s at 500-time units and beyond, indicating complete thermalization of the fluid. The inset shows that the amplitude of the thermalized $E(k,t)$ increases with time.}
   \label{fig:fig_power_spectrum}
\end{figure}

\begin{figure}[!ht]
    \centering
    \includegraphics[width=\columnwidth]{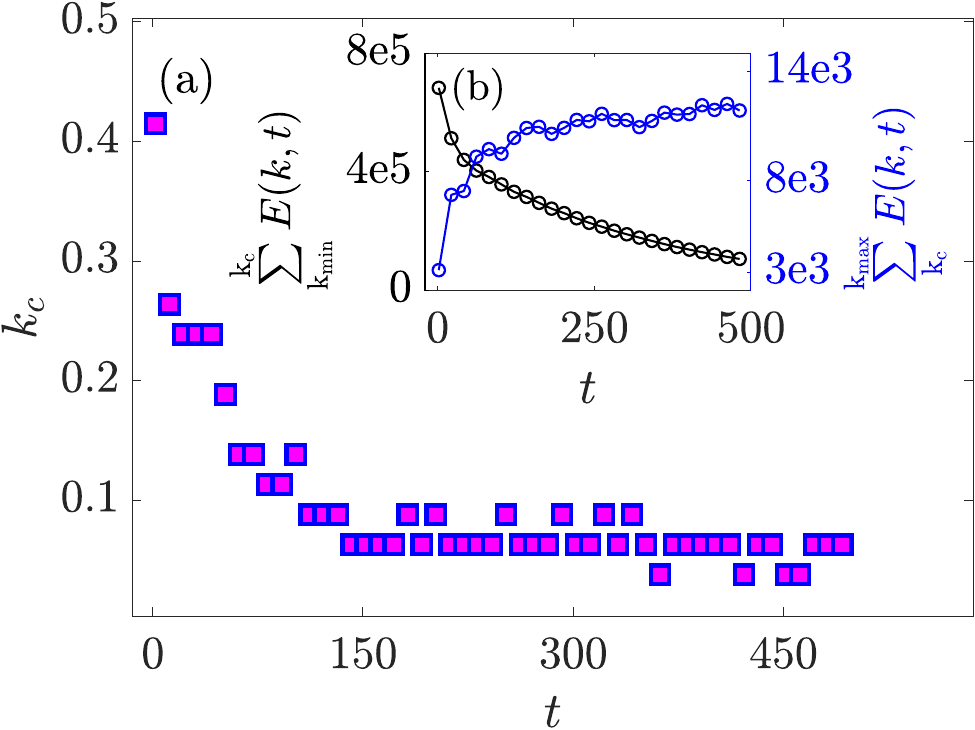}
    \caption{The time evolution of the transition wavevector $k_c$ that separates the coherent hydrodynamic energy (with $k^{-3}$ spectrum) and incoherent hydrodynamic energy (with $k$ spectrum). Note that $k_c$ decreases with time, as in Fig.~\ref{fig:fig_power_spectrum} (a) and (b). The inset exhibits a decrease in the total coherent energy $\sum_{k_\mathrm{min}}^{k_c} E(k,t)$ (black curve) and a growth in the total incoherent energy $\sum^{k_\mathrm{max}}_{k_c} E(k,t)$ (blue curve).}
    \label{fig:kc_evolution}
\end{figure}

Figure~\ref{fig:fig_power_spectrum}(a,b)  illustrates the energy spectra at various times ranging from $t =10$ to $2000$ LJ units. In these plots, the smallest wavenumber corresponding to the large-scale vortex is $k_0=2\pi/L \approx 0.0125$, whereas $k_\mathrm{max} = 40 k_0 \approx 0.5$. During the initial evolution, Fig.~\ref{fig:fig_power_spectrum}(a) exhibits $E(k,t) \propto k^{-3}$ for small wavenumbers, and $E(k,t) \propto k$ at large wavenumbers. The $k^{-3}$ energy spectrum is related to the vortex structure (as explained later), whereas the  $k$ spectrum represents the thermalization of the system at large wavenumbers. At around $t=50$, $E(k,t)$ at small $k$'s starts to deviate from $k^{-3}$ scaling because of the energy transfers from the vortex to the fluctuations. Note that $E(k)$ with spectral exponents between -3 and 1 represents broken hydrodynamic structures. Here,  the system exhibits nonequilibrium behavior at large scales and equilibrium features at small scales.
\\
As shown in Fig.~\ref{fig:fig_power_spectrum}(a,b), the regime of $k$ spectrum increases with time, and the system is fully thermalized $(E(k,t) \propto k$) at $t=500$ (or $\sim 14$ eddy turnover time) and beyond, which is consistent with the thermalization time for 3D Euler turbulence~\citep{Cichowlas_PRL_2005}. In addition, the magnitude of $E(k,t)$ decreases for small and intermediate wavenumbers, but it increases for large wavenumbers, similar to the evolution in 3D Euler turbulence~\cite{Cichowlas_PRL_2005} (refer to the inset of Fig.~\ref{fig:fig_power_spectrum}(b)). We denote $k_c$ as the transition wavenumber from where   $E(k) \propto k$, and quantify the coherent and incoherent energies of the hydrodynamic flow using   $\sum_{k_\mathrm{min}}^{k_c} E(k,t)$ and  $\sum^{k_\mathrm{max}}_{k_c} E(k,t)$ respectively. We illustrate these quantities in Fig.~\ref{fig:kc_evolution}. Clearly, $k_c$ decreases with time because of the expansion of the $k$ spectrum.  At $t=500$ LJ units, the coherent energy is nearly over, indicating complete thermalization of the system. Note that the particle velocity follows the Maxwell-Boltzmann distribution; thus, local thermal equilibrium is always maintained. This feature deviates from  Euler turbulence, which lacks a thermal component.  Using the Green-Kubo relation \cite{Kirova_2015}, we estimate the kinematic (shear) viscosity of the system to be $\nu = 8.7$ LJ units at temperature $T=100$. With this, the Reynolds number $\mathrm{Re} =UL/\nu \approx 812$.
\\
We have also determined the viscous relaxation time ($\tau$) by using the stress relaxation function, which is the auto-correlation of the shear stress tensor represented by $G(t) = (\langle \sigma_{xy}(t)\sigma_{xy}(0)\rangle)/(Ak_B T)$~\citep{AshwinJoy_PRL_2005}, where $A$ is the area system. The initial value of this auto-correlation yields the infinite frequency shear modulus $G_{\infty} = G(0)$. Therefore, the viscous relaxation time can be expressed as follows:
$\tau = [\int_{0}^{\infty} G(t) dt]/ (G_{\infty})$
We found the viscous relaxation time $\tau = 28$ LJ time units. All physical parameters are the same as in the rest of the paper.
The typical hydrodynamic vortex decay time is found to be 400 LJ-time units.
\\
\subsection{Energy and enstrophy fluxes}
\begin{figure}[!ht]
    \centering
    \includegraphics[width=\columnwidth]{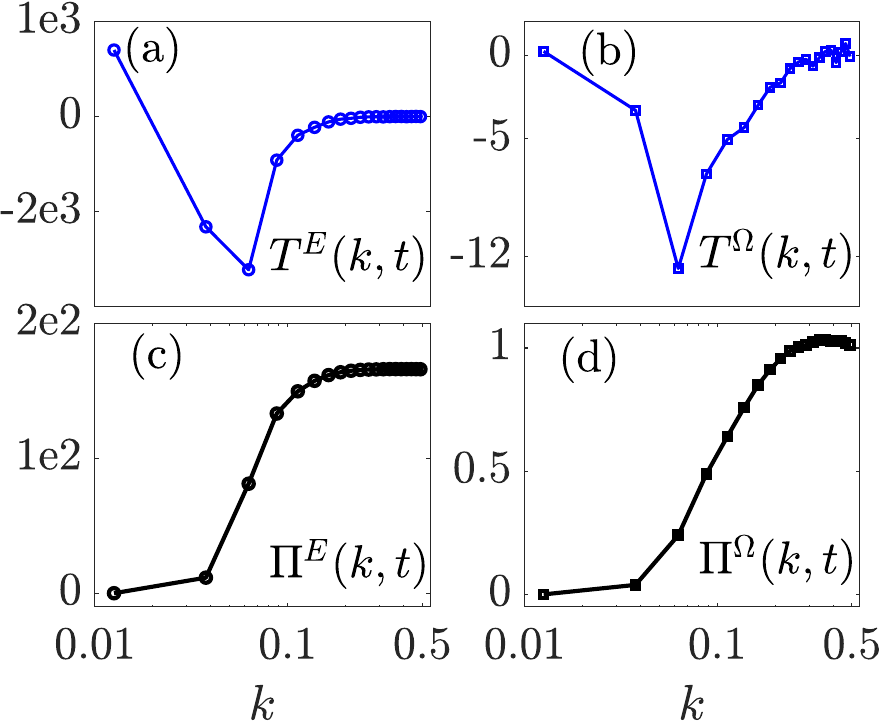}
    \caption{(a) Nonlinear energy transfer $T^E(k,t)$ and (b) nonlinear enstrophy transfer $T^\Omega(k,t) = k^2 T^E(k,t)$.  (c,d) The energy flux $\Pi^E(k,t)$ and the enstrophy flux $\Pi^E(k,t)$. }
    \label{fig:flux_nlt}
\end{figure}

Next, we compute the energy and enstrophy transfers and the corresponding fluxes of our flow. Using the energy equation we derive that $dE(k,t)/dt = T^E(k,t)$, where $T^E(k,t) = \sum_{\bf p}\Im[\{{\bf k}\cdot {\bf u(k-p},t)\}\{{\bf u(p},t) \cdot {\bf u^*(k},t)\}]$ is the nonlinear energy transfer term, and that the energy flux $\Pi^E(k) = -\int_{0}^{k} T^{E}(k') dk'$~\cite{Dar:PD2001,Verma:book:ET}. In addition,   $dE^\Omega(k,t)/dt = T^\Omega(k,t) = k^2 T^E(k,t)$, and the enstrophy flux $\Pi^{\Omega}(k,t) = -\int_{0}^{k} T^{\Omega}(k',t) dk' = -\int_{0}^{k} k'^2 T^{E}(k',t) dk'$~\cite{Verma:book:ET}.  Using $\int_{0}^{k} k'^2 |T^{E}(k',t)| dk' \le k^2  \int_{0}^{k}  |T^{E}(k',t)| dk'$, we deduce that $\Pi^\Omega(k,t) < k^2 \Pi^E(k,t)$ when $T^E(k,t)$ and $T^\Omega(k,t)$ are negative definite.
\\
We  compute $T^{E}(k,t)$, $T^{\Omega}(k,t)$, $\Pi^{E}(k,t)$, and $\Pi^{\Omega}(k,t)$ for our system. We estimate $T^{E}(k,t=17.5) \approx [E(k,t=25)-E(k,t=10)]/(2 \times 7.5)$ with an optimum value of $\Delta t = 7.5$ LJ units or 0.2 eddy turnover time. The above $\Delta t$ appears large, but it is reasonable considering large fluctuations in MD simulations. The quantities  $T^{E}(k,t)$, $T^{\Omega}(k,t)$, $\Pi^{E}(k,t)$,  $\Pi^{\Omega}(k,t)$ plotted in  Fig.~\ref{fig:flux_nlt} illustrate that $T^{E}(k,t), T^{\Omega}(k,t) <0$, indicating energy and enstrophy transfers from smaller wavenumbers to larger wavenumbers. This feature leads to positive energy and enstrophy fluxes, with $\Pi^{E,\mathrm{max}} \approx 166$ units, and $\Pi^{\Omega,\mathrm{max}} \approx 1$ unit.  These fluxes are consistent with the fact that $\Pi^\Omega(k,t) < k^2 \Pi^E(k,t)$ with $k \approx 0.1$.

Note that our MD simulation has an interesting feature that deviates from Euler turbulence. For Euler turbulence, $\int_0^\infty  T^E(k',t)dk' = 0$ and $\Pi^{E}(k,t) \rightarrow 0$ as $k \rightarrow \infty$. However, for MD simulations, $\int_0^\infty  T^E(k',t)dk' < 0$ and $\Pi^{E}(k,t)$ is finite at $k = k_\mathrm{max}$. This is because, in MD simulations, the grid-level hydrodynamic modes transfer energy to the thermal particles; this feature is absent in Euler turbulence, which has no thermal particles. 

\subsection{Energy spectrum at the early phase}
Lastly, we discuss the $k^{-3}$ energy spectrum exhibited in Fig.~\ref{fig:fig_power_spectrum}(a) during the early phase of our simulation.\citet{Kraichnan:JFM1973} proposed that 2D hydrodynamic turbulence forced at wavenumber $k_f$ exhibits dual cascade. The kinetic energy cascades to small wavenumbers (large length scales), leading to $k^{-5/3}$ energy spectrum, whereas the enstrophy exhibits a forward cascade leading to $k^{-3}$ energy spectrum.  Our simulation results are in agreement with Kraichnan’s theory. Note that our MD simulation is not forced. Therefore, the large-scale structures effectively feed energy at the largest scale, which is the box size. Hence, the kinetic energy cannot flow backward, thus ruling out the $k^{-5/3}$ energy spectrum. Instead, the enstrophy cascades forward from a small wavenumber, yielding $k^{-3}$ energy spectrum before the coherent energy starts to decay. In the $k^{-3}$ spectral regime, $T^\Omega (k,t) = k^2 T^E(k,t)$ that leads to a dominance of kinetic energy flux ($\Pi^E$) compared to the enstrophy flux ($\Pi^\Omega$). This is observed in our simulation.

\begin{figure}[!ht]
    \centering
    \includegraphics[width=\columnwidth]{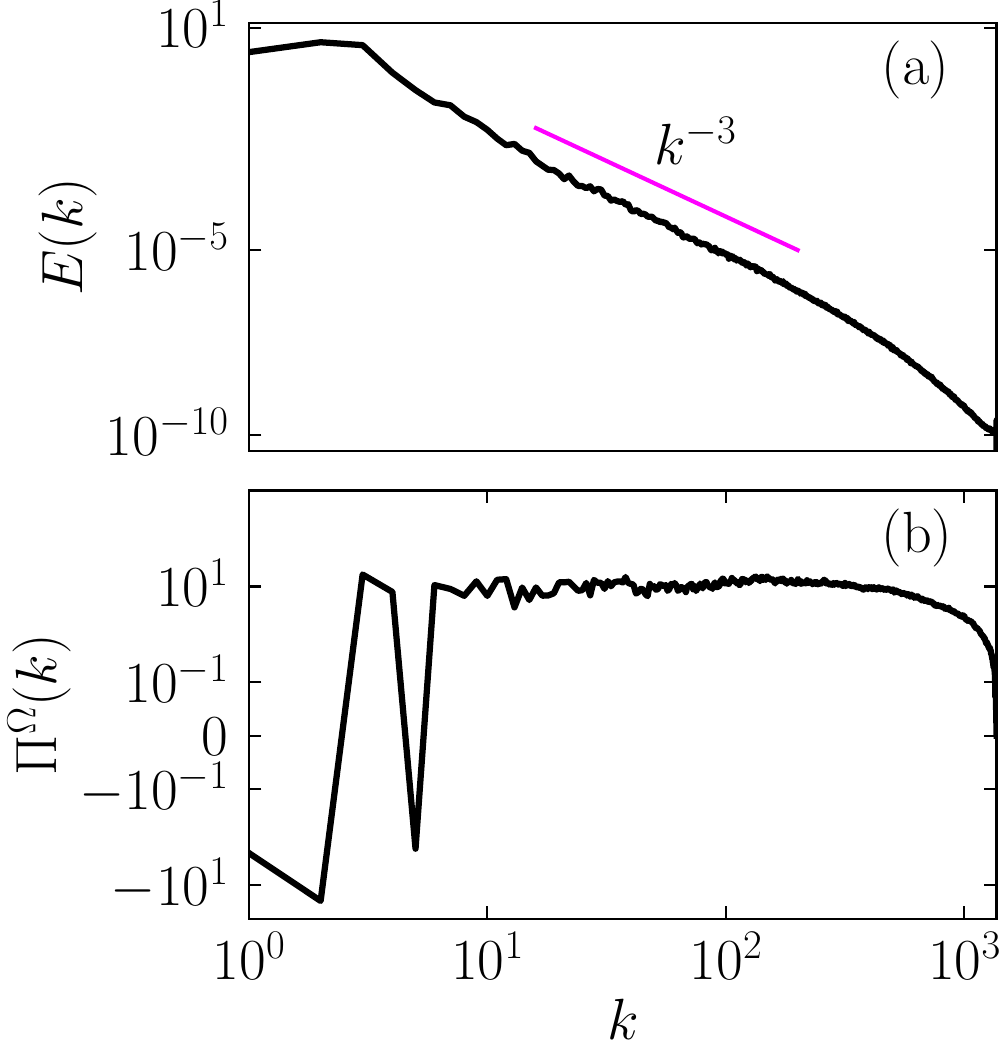}
    \caption{(a) Energy spectra obtained from Direct Numerical Simulations (DNS) exhibiting $k^{-3}$ spectrum in the inertial range, (b) The enstrophy flux $\Pi^\Omega(k)$, which is constant in this range.}
    \label{fig:DNS}
\end{figure}

Several hydrodynamic simulations exhibit $k^{-3}$ spectrum at large scales.  We performed a direct numerical simulation of 2D decaying hydrodynamics (Navier-Stokes equation) in a periodic box of size $(2\pi)^2$ and $4096^2$ grid using a pseudospectral code TARANG~\citep{verma2013benchmarking}. We chose kinematic viscosity $\nu = 7\times 10^{-6}$ and initial condition of Eq.~(2) with $k_0=1$. Note that the spectral simulation has four vortices due to the Fourier basis. After around an eddy turnover time, the system evolves to a state with $k^{-3}$ energy spectrum and a constant enstrophy flux. Refer to Fig.~\ref{fig:DNS} for an illustration. Note that the structures decay in our simulation due to lack of forcing.

The $k^{-3}$ spectrum described above is related to the similar spectrum in Earth’s atmosphere, which is quasi-two-dimensional. \citet{Nastrom:Nature1984} and \citet{Gage:JAS1986} measured the velocity field of the atmosphere and reported $k^{-3}$ energy spectrum at the synoptic scale from 500 to 3000 km) and $k^{-5/3}$ energy spectrum thereafter~\citet{Skamarock_JAS_2014}. Several numerical simulations of Earth's atmosphere, e.g., \citet{Skamarock_JAS_2014}, \citet{Wang_JAS_2021} reported similar spectra. Researchers have provided wide-ranging explanations for these phenomena~\citep{lindborg1999can}. We believe that our present results are related to  Earth's atmospheric turbulence. In literature~\citep{Skamarock_JAS_2014,Wang_JAS_2021}, it has been argued that baroclinic waves force the atmosphere at the synoptic scale, which is the largest scale of the system. Following the arguments stated above, we may argue that the large-scale structures of the Earth's atmosphere exhibit $k^{-3}$ energy spectrum due to the enstrophy cascade. We plan to test these arguments using detailed numerical simulations and atmospheric data at a later date.

Interestingly, there is another way to derive $k^{-3}$ energy spectrum for a smooth vortical flow structure. For such flows,  the two-point correlation function $C({\bf r},t) = \langle {\bf u(x+r},t)  {\bf u(x},t) \rangle \propto \alpha - \beta r^2 $, with $\alpha$ and $\beta$ as constants. Therefore, $C({\bf k},t) = \int d{\bf r} C({\bf r},t) \exp(i {\bf k \cdot r}) \propto k^{-4}$, and hence $E(k,t) = C({\bf k},t)2\pi k \sim k^{-3}$.
\section{Summary}
Finally, we summarize our results. We study hydrodynamic features such as turbulent energy flux and thermalization in a many-particle system. We isolate our system from the heat bath so as to conserve the total energy and start it with a hydrodynamic vortex. In the early phase, the system exhibits $E(k) \propto k^{-3}$ at small wavenumbers and $E(k) \propto k$ at large wavenumbers. The system gets fully thermalized at $t \gtrsim 500$ LJ units, at which point the coherent hydrodynamic energy is fully transferred to the incoherent hydrodynamic component ($E(k) \propto k$) and thermal noise. We strengthen our arguments using the energy and enstrophy transfers. Our MD simulation clearly demonstrates how large-scale kinetic energy cascades to small scales in a realistic turbulent system.   

Our results are consistent with the route to thermalization observed earlier in Euler turbulence~\cite{Cichowlas_PRL_2005}, in Navier-Stokes equation with fluctuating hydrodynamics~\cite{Bell:JFM2022,Bandak:PRE2022}, and in DSMC simulations~\cite{McMullen_2022_PRL}. However, there are some subtle differences. In our simulation, some of the hydrodynamic energy is transferred to the thermal particles at the microscopic level, which is absent in 3D Euler turbulence. We also remark that particle dynamics is more realistic in MD simulations than in DSMC.  

In our system, the large-scale initial structure generates $k^{-3}$ energy spectrum, as well as forward energy and enstrophy fluxes.   However, the vortex structure in our system breaks down and thermalizes due to a lack of external forcing. We also note that our system exhibits near-incompressibility, with maximum and average density fluctuations of $0.15$ and $0.05$, respectively.
We also observe similar thermalization for other initial conditions, such as sheared flow, wave excitation, Rayleigh-Taylor instability, and four-vortex during the validation process. Interestingly, in earlier simulations of 2D Euler systems with a set of vortices as initial conditions are out of equilibrium~\cite{Bouchet:PRL2009,Bouchet:PR2012,Verma:PRF2022}.  Resolution of  this issue needs further investigation.

Thus, our MD simulations of hydrodynamic structures embedded in a noisy environment provide valuable insights into the thermalization process in a realistic fluid system. The work also sheds light on the connection between hydrodynamics and thermodynamics, which are active at macroscopic and microscopic scales, respectively.

\section{Acknowledgements}
The author thanks J\"org Schumacher, Rodion Stepanov, and Soumyadeep Chatterjee for valuable comments. The authors acknowledge the use of AGASTYA HPC for present studies. 
RW and ST also acknowledge the support for this work through SERB Grant No. CRG/2020/003653, and MKV, the support of SERB Grant Nos. SERB/PHY/20215225 and SERB/PHY/2021473.


%

\end{document}